\documentclass[preprint,12pt]{elsarticle}

\usepackage{graphicx}
\usepackage{amssymb}
\usepackage{amsthm}
\usepackage{amsmath}

\usepackage{lineno}

\usepackage{siunitx}

\journal{Nuclear Instruments and Methods in Physics Research A}

\begin{document}

\begin{frontmatter}

%% Title, authors and addresses

\title{Wigner-Eckart Theorem and the False EDM of $^{199}$Hg}

%% use the tnoteref command within \title for footnotes;
%% use the tnotetext command for the associated footnote;
%% use the fnref command within \author or \address for footnotes;
%% use the fntext command for the associated footnote;
%% use the corref command within \author for corresponding author footnotes;
%% use the cortext command for the associated footnote;
%% use the ead command for the email address,
%% and the form \ead[url] for the home page:
%%
%% \title{Title\tnoteref{label1}}
%% \tnotetext[label1]{}
%% \author{Name\corref{cor1}\fnref{label2}}
%% \ead{email address}
%% \ead[url]{home page}
%% \fntext[label2]{}
%% \cortext[cor1]{}
%% \address{Address\fnref{label3}}
%% \fntext[label3]{}

%% use optional labels to link authors explicitly to addresses:
%% \author[label1,label2]{<author name>}
%% \address[label1]{<address>}
%% \address[label2]{<address>}

\author[um]{W.~Klassen}
\author[uw,um]{J.W.~Martin}
\author[lspc]{G.~Pignol}

\address[um]{Physics and Astronomy, University of Manitoba, Winnipeg, MB, Canada}
\address[uw]{Department of Physics, The University of Winnipeg, Winnipeg, MB, Canada}
\address[lspc]{Universit\'e Grenoble Alpes, CNRS, Grenoble INP, LPSC-IN2P3, Grenoble, France}

\begin{abstract}
  In neutron electric dipole moment (EDM) experiments, $^{199}$Hg is used as
  a comagnetometer.  The comagnetometer suffers from a false EDM
  arising in leading order from a gradient $\partial B_{z}/\partial z$
  in the magnetic field.  Our work concerns higher-order multipole
  corrections to the false EDM of $^{199}$Hg.  We show that for
  spherical traps, all higher-order multipoles are identically zero.
  We further show that for the usual cylindrical traps used in EDM
  experiments, selection of quasi-spherical dimensions for the trap
  can reduce the higher-order contributions.  The results are another
  indication that trap geometry is an important consideration for
  experiments desiring to control this systematic effect.
\end{abstract}

\begin{keyword}
false electric dipole moment \sep trapped particles \sep neutron electric dipole moment
%% keywords here, in the form: keyword \sep keyword

%% MSC codes here, in the form: \MSC code \sep code
%% or \MSC[2008] code \sep code (2000 is the default)

\end{keyword}

\end{frontmatter}

%%
%% Start line numbering here if you want
%%
%\linenumbers

%% main text
\section{False electric dipole moments for particles in traps}
\label{sec:intro}

In the most precise neutron electric dipole moment (EDM) experiments,
ultracold neutrons are stored in a bottle in either parallel or
antiparallel electric $E$ and magnetic $B$ fields.  Their spin
precession frequency $\omega_\pm$ is measured
\begin{equation}
\omega_\pm=(2\mu_nB\pm 2d_nE)/\hbar
\end{equation}
for each of the parallel ($+$) and antiparallel ($-$) configurations,
leading to a determination of the neutron electric dipole moment
$d_n$.  Here $\mu_n$ is the neutron magnetic moment.

A crucial aspect of the experiment is that the magnetic field be
continuously monitored so that any drifts can be corrected.  To
monitor the field, a second ``comagnetometer'' atomic species is
stored in the cell and its spin precession frequency is measured
optically.  Normally $^{199}$Hg is used for this
purpose~\cite{bib:green,bib:bakernim,bib:hg-better}, in part because
its true EDM has been constrained to be
small~\cite{bib:griffith,bib:hgupdate}.

Ideally, the magnetic field $B$ should be uniform so that long
free-precession times can be achieved for both the neutrons and Hg
atoms.  In the previous most precise nEDM
experiment~\cite{bib:baker,bib:pendlebury} it was found that a
vertical gradient $\partial B_z/\partial z$ in the magnetic field
induced false EDM's for the neutrons and Hg atoms.  The frequency
shift can be considered as a Ramsey-Bloch-Siegert shift for particles
traveling in orbits within the trap~\cite{bib:gpe1,bib:gpe2}.  The
ultracold neutrons traverse the measurement volume slowly enough that
their accrued phase can be thought of as a geometric
phase which accrues adiabatically~\cite{bib:gpe1}.

The false EDM's can also be calculated by correlation function
techniques~\cite{bib:gpe3,bib:gpe4}.  This led to the realization that
false EDM for the Hg atoms could be written in the form
\begin{equation}
\label{eq:pignol}
  d_{f,\rm Hg}=-\frac{\hbar\gamma^{2}}{2c^{2}}\langle xB_x+yB_y\rangle
\end{equation}

using integration by parts~\cite{bib:pignol-roccia}.  Here, $\gamma=2\mu/\hbar$
is the gyromagnetic ratio of Hg with $\mu$ being its magnetic moment, and the average is over the
storage volume. This form is valid to high precision in the
low-frequency (field) limit, even reproducing higher-order effects
first studied using Monte Carlo techniques in Ref.~\cite{bib:gpe2}.

The false EDM of Hg is also well-understood experimentally, having
been characterized using surrounding Cs magnetometers~\cite{bib:hgcs}
and to higher orders~\cite{bib:komposch}.

The work presented here concerns higher-order corrections beyond the
first-order vertical gradient $\partial B_{z}/\partial z$.  We show
that for spherical traps, all higher-order terms contributing to
Eq.~(\ref{eq:pignol}) are identically zero.  We further show that for
the usual cylindrical traps used in EDM experiments, selection of
quasi-spherical dimensions for the trap can reduce the higher-order
contributions to false EDM of the Hg atoms.

\section{Harmonic Decomposition of the Magnetic Field}

Within the measurement region of EDM experiments, nonmagnetic
components are used so that the field can be measured and controlled
precisely.  Since bound and free currents are absent from this region,
the magnetic field can be written as
\begin{equation}
\vec{B}=-\nabla\Phi_M
\end{equation}
where $\Phi_M$ is the magnetic scalar potential.  Since
$\nabla\cdot\vec{B}=0$, $\Phi_M$ obeys Laplace's equation:
\begin{equation}
\nabla^2\Phi_M=0.
\end{equation}
The general solution for a boundary-value problem can be written in
spherical coordinates $(r,\theta,\phi)$ in terms of spherical
harmonics $Y_{\ell m}(\theta,\phi)$ as~\cite{bib:jackson}
\begin{equation}\label{eq:boundaryvalue}
  \Phi_M(r,\theta,\phi)=\sum_{\ell=0}^\infty\sum_{m=-\ell}^\ell\left[A_{\ell m}r^\ell+B_{\ell m}r^{-(\ell+1)}\right]Y_{\ell m}(\theta,\phi)
\end{equation}
where $A_{\ell m}$ and $B_{\ell m}$ are sets of constants determined
by the boundary conditions.  If we define $r=0$ to be the center of
the trap, the requirement that $\vec{B}$ remain finite enforces
$B_{\ell m}=0$.  The spherical harmonics can be written as
\begin{equation}
    Y_{\ell m}(\theta,\phi)=\sqrt{\frac{2\ell+1}{4\pi}\frac{(\ell-m)!}{(\ell+m)!}}P_{\ell}^{m}(\cos\theta)e^{im\phi},
\end{equation}
where $P_{\ell}^{m}$ are the associated Legendre polynomials.  

The average appearing in Eq.~(\ref{eq:pignol}) can be recast in terms
of the scalar potential as
\begin{equation}
  \langle xB_x+yB_y\rangle=-\left\langle\left(x\frac{\partial}{\partial
    x}+y\frac{\partial}{\partial y}\right)\Phi_M\right\rangle
\end{equation}
which in terms of the spherical harmonics becomes
\begin{equation}
  \label{eq:xb}
  \langle xB_x+yB_y\rangle=-\sum_{\ell=0}^\infty\sum_{m=-\ell}^\ell
  A_{\ell m}\left\langle\left(x\frac{\partial}{\partial
    x}+y\frac{\partial}{\partial y}\right)r^\ell Y_{\ell
    m}\right\rangle
\end{equation}
where again the average is conducted over the measurement cell.

\section{Application of the Wigner-Eckart Theorem}

We can apply the Wigner-Eckart Theorem to this system by making an
analogy to matrix elements in quantum mechanics.  The differential
operator in Eq.~(\ref{eq:xb}) $\left(x\frac{\partial}{\partial
  x}+y\frac{\partial}{\partial y}\right)$ can be analogized to the
operator $(xp_x+yp_y)$ in quantum mechanics.

For a particle in a spherically symmetric potential, the stationary
states may be written as eigenstates of the angular momentum operators
\begin{align}
L^2|n\ell m\rangle&=\hbar^2\ell(\ell+1)|n\ell m\rangle\\
L_z|n\ell m\rangle&=\hbar m|n\ell m\rangle
\end{align}
where $|n\ell m\rangle$ represents the stationary state and $\ell$ and
$m$ are quantum numbers.  The quantum number $n$ would count energy
levels.  In the position representation, the states factorize in
spherical coordinates as
\begin{equation}
\langle\vec{r}|n\ell m\rangle=R_{n\ell}(r)Y_{\ell m}(\theta,\phi)
\end{equation}
and thus the states are related to the spherical harmonics.

Eq.~(\ref{eq:xb}) therefore has a number of elements which bear a
strong similarity with calculations in quantum mechanics.  The average
in Eq.~(\ref{eq:xb}) is over the EDM measurement cell volume.  In
order to use the Wigner-Eckart theorem, we want to make an analogy to
quantum mechanical matrix elements.  In order for this analogy be
valid, the cell would have to be a spherical cell.  In this
circumstance, the term under the average is proportional to a quantum
mechanical matrix element as
\begin{equation}
\label{eq:matrix}
  \left\langle\left(x\frac{\partial}{\partial
  x}+y\frac{\partial}{\partial y}\right)r^\ell Y_{\ell
  m}\right\rangle\sim\langle n'00|(xp_x+yp_y)|n\ell m\rangle
\end{equation}
where we have inserted a spherically symmetric state with
$\ell'=m'=0$.  Being constructed from products of vector operators, we
will show that the operator $(xp_x+yp_y)$ can be written as an
admixture of spherical tensors of rank 0 and 2.

The Wigner-Eckart theorem relates matrix elements of spherical tensors
involving these states to Clebsch-Gordan coefficients
\cite{bib:sakurai}.  Applied to two states $|n'\ell'm'\rangle$ and
$|n\ell m\rangle$, it would read
\begin{equation}
\langle n'\ell'm'|T_q^{(k)}|n\ell m\rangle=\langle\ell k;mq|\ell k;\ell'm'\rangle\frac{\langle n'\ell'||T^{(k)}||n\ell\rangle}{\sqrt{2j+1}}
\end{equation}
where $T_q^{(k)}$ is a spherical tensor of rank $k$ with magnetic
quantum number $q$.  The double-bar matrix element is a reduced matrix
element that does not depend on $m$, $m'$, or $q$.

The Clebsch-Gordan coefficient $\langle\ell k;mq|\ell
k;\ell'm'\rangle$ can be thought of in the following more familiar
way.  Imagine two angular momentum operators $\vec{L}$ and $\vec{K}$
with simultaneous eigenstates $|\ell k;mq\rangle$ s.t.
\begin{align}
L^2|\ell k;mq\rangle&=\hbar^2\ell(\ell+1)|\ell k;mq\rangle\\
L_z|\ell k;mq\rangle&=\hbar m|\ell k;mq\rangle\\
K^2|\ell k;mq\rangle&=\hbar^2k(k+1)|\ell k;mq\rangle\\
K_z|\ell k;mq\rangle&=\hbar q|\ell k;mq\rangle.
\end{align}
If we now define a new operator $\vec{L}'=\vec{L}+\vec{K}$, the
Clebsch-Gordan coefficients represent the coefficients transforming to
the new basis $|\ell k;\ell'm'\rangle$ where
\begin{align}
L^2|\ell k;mq\rangle&=\hbar^2\ell(\ell+1)|\ell k;\ell'm'\rangle\\
K^2|\ell k;mq\rangle&=\hbar^2k(k+1)|\ell k;\ell'm'\rangle\\
L'^2|\ell k;mq\rangle&=\hbar^2\ell'(\ell'+1)|\ell k;\ell'm'\rangle\\
L'_z|\ell k;mq\rangle&=\hbar m'|\ell k;\ell'm'\rangle.
\end{align}
In this way, the Clebsch-Gordan coefficient is related to the addition
of angular momentum of the state $|n\ell m\rangle$ to that of the
spherical tensor $T_q^{(k)}$ and reaching the state
$|n'\ell'm'\rangle$.

The Clebsch-Gordan coefficient is only non-zero if
\begin{equation}
  \label{eq:z}
  m+q=m'
\end{equation}
and
\begin{equation}
\label{eq:ell}
  |\ell-k|\leq\ell'\leq\ell+k
\end{equation}

As mentioned earlier the operator $(xp_x+yp_y)$ can be written as an
admixture of spherical tensors of rank 0 and 2.  The following two
spherical tensors may be constructed from the Cartesian components of
the vector operators $\vec{r}$ and $\vec{p}$:
\begin{equation}
T_0^{(0)}=-\frac{\vec{r}\cdot\vec{p}}{3}=\frac{-xp_x-yp_y-zp_z}{3}
\end{equation}
and
\begin{equation}
T_0^{(2)}=\frac{3zp_z-\vec{r}\cdot\vec{p}}{\sqrt{6}}=\frac{-xp_x-yp_y+2zp_z}{\sqrt{6}}.
\end{equation}
The operator of interest can then be written as
\begin{equation}
xp_x+yp_y=-2T_0^{(0)}-\sqrt{\frac{2}{3}}T_0^{(2)}
\end{equation}
which are spherical tensors of rank 0 and 2 with magnetic quantum
number $q=0$.

Finally we note that because these operators are even under the parity
transformation, they may only link states of the same parity.

We can now apply Eqs.~(\ref{eq:z}) and (\ref{eq:ell}) to
Eq.~(\ref{eq:xb}) to find that only terms in the sums with $m=0$ and
$\ell=0,1,2$ will contribute.  Since the operator in Eq.~(\ref{eq:xb})
is a differential operator, the term with $\ell=0$ also cannot
contribute.  The term with $\ell=1$ is ruled out by the parity
selection rule.  We are then left with only one non-zero term arising
from the harmonic decomposition of the field:
\begin{equation}
\langle
xB_x+yB_y\rangle=-A_{20}\left\langle\left(x\frac{\partial}{\partial
  x}+y\frac{\partial}{\partial y}\right)r^2 Y_{20}\right\rangle
\end{equation}
where the average can be readily carried out since $r^2Y_{20}$ is a
polynomial of degree 2 in Cartesian coordinates.

The suppression of the higher-order terms can be derived in a number
of different ways, for example, by the commutation relations of the
operators, by integration by parts, and/or by using the known
properties of the spherical harmonics/Legendre polynomials.  We used
this to check the result in a number of ways.  This included (a)
explicit integration of the $m=0$ terms for particular higher $\ell$,
demonstrating they were all zero, and (b) integration by parts
(shifting the derivatives to the left) and then using the
orthogonality of the Legendre polynomials to demonstrate that only
$\ell=2$ is permitted.

\section{Cylindrical Trap and Suppression of Higher Orders}
\label{sec:cylinder}

A requirement of the calculation of the preceding section is that the
EDM cell be spherical, which is not an attractive option for EDM
experiments.  A more typical geometry is a cylindrically symmetric
geometry.  In this case, it can still readily be demonstrated that
only $m=0$ terms contribute in the second sum in Eq.~(\ref{eq:xb}).
Switching to cylindrical $(\rho,\phi,z)$ coordinates, the differential
operator becomes $x\frac{\partial}{\partial
  x}+y\frac{\partial}{\partial y}=\rho\frac{\partial}{\partial\rho}$
and since $Y_{\ell m}\sim e^{im\phi}$ these terms would average to
zero unless $m=0$.  Furthermore, terms with $\ell={\rm odd}$ are odd
in $z$.  Thus if the cylindrical cell is centered on $z=0$, these
terms will also average to zero.  We are then left with
\begin{equation}
  \label{eq:cyl}
  \langle xB_x+yB_y\rangle=-\sum_{\ell=2,4,6,...}^\infty A_{\ell
    0}\left\langle\left(\rho\frac{\partial}{\partial
    \rho}\right)r^\ell Y_{\ell
    0}\right\rangle
\end{equation}
The purpose of this section is to demonstrate that $\ell=2$ dominates,
that for a certain choice of cell dimensions the $\ell=4$ term can be
zeroed, and that for this selection the $\ell=6$ term can be reduced
compared to the typical cell geometry used for the ILL nEDM
experiment.

In order to more easily compare to Ref.~\cite{bib:pignol-roccia}, we
introduce a more convenient normalization based on the Legendre
polynomials $P_\ell(\cos\theta)$ as
\begin{equation}
\label{eq:cylprime}
  \langle xB_x+yB_y\rangle= +
    \sum_{\ell=2,4,6...}^{\infty}\frac{g_{(\ell-1)0}}{\ell}\left\langle\left(\rho\frac{\partial}{\partial
      \rho}\right)r^\ell P_\ell(\cos\theta)\right\rangle,
\end{equation}
where the $g_{(\ell-1)0}$ are related by constants (and a sign change)
to the $A_{\ell 0}$ in Eq.~(\ref{eq:cyl}).  This normalization is
preferred because it ensures that the $m=0$ components of
$B_{z}=-\frac{\partial\Phi_M}{\partial z}$ are polynomials which
contain $g_{\ell 0}z^\ell$ with $\ell={\rm odd}$.  This guarantees,
for example, that the leading-order term in the vertical gradient is
$\frac{\partial B_z}{\partial z}=g_{10}$, allowing us to easily
identify $g_{10}$ itself as the first-order uniform gradient term of
Ref.~\cite{bib:pignol-roccia}.

%The Legendre polynomials of interest are
%\begin{equation}
%P_{2}^{0}(\cos\theta)=\frac{1}{2}(3\cos^{2}\theta-1)
%\end{equation}
%\begin{equation}
%P_{4}^{0}(\cos\theta)=\frac{1}{8}(35\cos^{4}\theta-30\cos^{2}\theta+3)
%\end{equation}
%and
%\begin{equation}
%P_{6}^{0}(\cos\theta)=\frac{1}{16}(231\cos^{6}\theta-315\cos^{4}\theta+105\cos^{2}\theta-5).
%\end{equation}

In general $\frac{1}{\ell}r^\ell P_\ell(\cos\theta)$, when expressed
in cylindrical coordinates, is a polynomial in $\rho$ and $z$.  For
the terms of interest, the polynomials are
\begin{align}
\frac{1}{2}r^2P_2(\cos\theta)&=\frac{1}{4}\left(2z^2-\rho^2\right)\\
\frac{1}{4}r^4P_4(\cos\theta)&=\frac{1}{32}\left(8z^{4}-24z^{2}\rho^{2}+3\rho^{4}\right),~{\rm and}\\
\frac{1}{6}r^6P_6(\cos\theta)&=\frac{1}{96}\left(16z^{6}-120z^{4}\rho^{2}+90z^{2}\rho^{4}-5\rho^{6}\right).
\end{align}
We define the origin of coordinates to lie at the center of a
measurement cell with height $H$ and radius $R$.  Carrying out the
average over the cell, the first three non-zero terms of
Eq.~(\ref{eq:cylprime}) become
\begin{multline}
\label{eq:xbc}
  \langle xB_x+yB_y\rangle
=-\frac{R^{2}}{4}\left[g_{10}
+g_{30}\left(\frac{H^{2}}{4}-\frac{R^2}{2}\right)\right.\\
  \left.+g_{50}\left(\frac{H^{4}}{16}-\frac{5R^2H^2}{12}+\frac{5R^4}{16}\right)\right].
\end{multline}
The expression for $d_{f,\rm Hg}$ then becomes
\begin{multline}
d_{f,\rm Hg}=\frac{\hbar\gamma^{2}R^2}{8c^{2}}\left[g_{10}
+g_{30}\left(\frac{H^{2}}{4}-\frac{R^2}{2}\right)\right.\\
  \left.+g_{50}\left(\frac{H^{4}}{16}-\frac{5R^2H^2}{12}+\frac{5R^4}{16}\right)\right].
\end{multline}
The $g_{10}$ term is in agreement with Ref.~\cite{bib:pignol-roccia},
and the other two terms have also been derived
previously~\cite{bib:pignol-priv}.  We now analyze these next two
terms.

%The traditional units used in EDM experiments is e$\cdot$cm, and
%gradients are usually expressed in units per distance to some power.
%We write the units of the pre-factor:
%\begin{equation}
%   blank
%\end{equation}

An immediate observation is that the $g_{30}$ term can be set to zero
if the measurement volume is quasi-spherical {\it i.e.} for
\begin{equation}
\label{eq:h}
H=\sqrt{2}R
\end{equation}
If this selection is made, then the factor in the $g_{50}$ term becomes
\begin{equation}
\frac{H^{4}}{16}-\frac{5R^2H^2}{12}+\frac{5R^4}{16}=-\frac{13}{48}R^4
\end{equation}
The question now becomes whether this is a reasonable geometry for an
nEDM experiment.

\section{Discussion and Caveats}

\subsection{ILL/PSI nEDM experiment geometry}

For the ILL/PSI nEDM experiment, the cell dimensions are $R=23.5$~cm
and $H=12$~cm, and so Eq.~(\ref{eq:h}) is clearly not obeyed.  Using
the ILL/PSI geometry, the $g_{30}$ term is
\begin{equation}
\frac{H^{2}}{4}-\frac{R^2}{2}=-240~\mathrm{cm}^2
\end{equation}
and the $g_{50}$ term is
\begin{equation}
\frac{H^{4}}{16}-\frac{5R^2H^2}{12}+\frac{5R^4}{16}=\num{6.35e4}~\mathrm{cm}^4
\end{equation}
The false EDM then becomes
\begin{equation}
\label{eq:falseill}
d_{f,\rm Hg} = \num{1.15e-27}\left(g_{10} - g_{30}\cdot240 +
g_{50}\cdot\num{6.35e4}\right)\mathrm{e}\cdot\mathrm{cm}
\end{equation}
where the $g_{\ell0}$ are expressed in units of pT/cm$^\ell$.  The
$g_{30}$ term was analyzed and measured experimentally in
Ref.~\cite{bib:komposch} and found to be in agreement with
expectation.

\subsection{Quasi-spherical geometry}

In order to compare this result for a realistic experimental geometry
to our suggestion of a quasi-spherical cell, we considered several
possibilities.  One possibility was to keep the radius of the cell the
same as for the ILL experiment.  This would mean that the size of the
pre-factor in Eq.~(\ref{eq:falseill}) would be equal.  In this case,
the height of the cell would necessarily be increased to
$H=\sqrt{2}\cdot$23.5~cm~$\approx 33$~cm.  The negative aspect of this
suggestion is that a high-voltage nearly a factor of three larger
would need to be sustained across the electrodes in order to keep the
electric field in the experiment the same.  Since the electric field
is strongly related to the statistical precision of the experiment, it
is not clear that this is a realistic compromise.

Another possibility was to keep the height of the cell the same, which
would alleviate any high-voltage issues.  In this case the cell radius
would need to shrink to $R=12$~cm$/\sqrt{2}\approx 8.5$~cm.  The
negative aspect here is that the volume of the experiment would shrink
by a factor of almost eight.  In general, the statistical precision of
neutron EDM experiments is driven by the neutron density achievable in
the cell, and this choice would mean that the number of neutrons
loaded into the cell would be reduced by the same factor.

We therefore decided to suggest keeping approximately the same EDM
cell volume, setting $R=17$~cm and $H=\sqrt{2}\cdot$17~cm~$\approx
24$~cm.  For this choice, the $\ell=4$ term is zero and the expression
for the false EDM becomes
\begin{equation}
d_{f,\rm Hg} = \num{6.1e-28}\left(g_{10} - g_{50}\cdot\num{2.26e4}\right)\mathrm{e}\cdot\mathrm{cm}.
\end{equation}
For this particular choice, the overall false EDM is smaller, and
furthermore the $g_{50}$ term is reduced relative to the $g_{10}$ when
compared to the usual ILL/PSI geometry.

\subsection{Further notes}

It is unclear whether this geometry would be realizable
experimentally.  The main point of the discussion is that the geometry
affects the size of the $g_{\rm 30}$ term for the Hg comagnetometer
false EDM, and the closer to spherical the less the higher-order terms
tend to contribute.

It is also important to consider the overall false-EDM correction
scheme used in these experiments.  While the false EDM of the mercury
comagnetometer tends to dominate the correction, it is by no means the
only quantity affected by the higher multipoles of the magnetic field.
The neutrons' false EDM, the spin relaxation times of the species, and
the electric field independent spin-precession frequencies are also
affected by the inhomogeneity.

In particular, the spin-precession frequency ratio (neutrons to Hg)
was used in the ILL nEDM experiment to sense the vertical gradient
$g_{10}$ and hence to correct the leading terms in the false
EDM's~\cite{bib:baker,bib:pendlebury}.  This works because the
ultracold neutrons tend to preferentially sample the bottom of the
cell, resulting in an average height difference against gravity.  When
considered to higher order, other terms contribute to this correction
scheme.  So while the false EDM of the Hg atoms could be suppressed by
a quasi-spherical cell, it is unclear that this is the most important
factor in experiment design.  Even if the EDM cell could be made
spherical, the neutrons still preferentially sample the lower portion
of the sphere.

\section{Conclusion}

We have demonstrated that a spherical EDM measurement cell would
reduce the false EDM of Hg to depend on a single multipole of the
magnetic field.  A cylindrical cell that is quasi-spherical with
$H=\sqrt{2}R$ also tends to reduce the higher multipole contributions.
Both suggestions are rather difficult to realize experimentally, and
do not capture the full correction scheme used in the ILL/PSI nEDM
experiment, which must take more into account than simply the false
EDM of the Hg comagnetometer.  Nonetheless, we think this is an
interesting result which points out the dependence of the false EDM
contributions on the measurement cell geometry.

\section*{Acknowledgements}

This work was undertaken, in part, thanks to funding from the Natural
Sciences and Engineering Research Council Canada, the Canada Research
Chairs program, and the Canada Foundation for Innovation.  G.P. is
supported by the ERC grant 716651-NEDM.

%%\section*{References}

\bibliography{wigner.bib}

\begin{thebibliography}{17}
\expandafter\ifx\csname natexlab\endcsname\relax\def\natexlab#1{#1}\fi
\providecommand{\url}[1]{\texttt{#1}}
\providecommand{\href}[2]{#2}
\providecommand{\path}[1]{#1}
\providecommand{\DOIprefix}{doi:}
\providecommand{\ArXivprefix}{arXiv:}
\providecommand{\URLprefix}{URL: }
\providecommand{\Pubmedprefix}{pmid:}
\providecommand{\doi}[1]{\href{http://dx.doi.org/#1}{\path{#1}}}
\providecommand{\Pubmed}[1]{\href{pmid:#1}{\path{#1}}}
\providecommand{\bibinfo}[2]{#2}
\ifx\xfnm\relax \def\xfnm[#1]{\unskip,\space#1}\fi
%Type = Article
\bibitem[{Green et~al.(1998)Green, Harris, Iaydjiev, May, Pendlebury, Smith,
  van~der Grinten, Geltenbort, and Ivanov}]{bib:green}
\bibinfo{author}{K.~Green}, \bibinfo{author}{P.~G. Harris},
  \bibinfo{author}{P.~Iaydjiev}, \bibinfo{author}{D.~J.~R. May},
  \bibinfo{author}{J.~M. Pendlebury}, \bibinfo{author}{K.~F. Smith},
  \bibinfo{author}{M.~van~der Grinten}, \bibinfo{author}{P.~Geltenbort},
  \bibinfo{author}{S.~Ivanov},
\newblock \bibinfo{title}{{Performance of an atomic mercury magnetometer in the
  neutron EDM experiment}},
\newblock \bibinfo{journal}{Nucl. Instrum. Meth.} \bibinfo{volume}{A404}
  (\bibinfo{year}{1998}) \bibinfo{pages}{381--393}.
%Type = Article
\bibitem[{Baker et~al.(2014)}]{bib:bakernim}
\bibinfo{author}{C.~A. Baker}, et~al.,
\newblock \bibinfo{title}{{Apparatus for Measurement of the Electric Dipole
  Moment of the Neutron using a Cohabiting Atomic-Mercury Magnetometer}},
\newblock \bibinfo{journal}{Nucl. Instrum. Meth.} \bibinfo{volume}{A736}
  (\bibinfo{year}{2014}) \bibinfo{pages}{184--203}.
%Type = Article
\bibitem[{Ban et~al.(2018)}]{bib:hg-better}
\bibinfo{author}{G.~Ban}, et~al.,
\newblock \bibinfo{title}{{Demonstration of sensitivity increase in mercury
  free-spin-precession magnetometers due to laser-based readout for neutron
  electric dipole moment searches}},
\newblock \bibinfo{journal}{Nucl. Instrum. Meth.} \bibinfo{volume}{A896}
  (\bibinfo{year}{2018}) \bibinfo{pages}{129--138}.
%Type = Article
\bibitem[{Griffith et~al.(2009)Griffith, Swallows, Loftus, Romalis, Heckel, and
  Fortson}]{bib:griffith}
\bibinfo{author}{W.~C. Griffith}, \bibinfo{author}{M.~D. Swallows},
  \bibinfo{author}{T.~H. Loftus}, \bibinfo{author}{M.~V. Romalis},
  \bibinfo{author}{B.~R. Heckel}, \bibinfo{author}{E.~N. Fortson},
\newblock \bibinfo{title}{{Improved Limit on the Permanent Electric Dipole
  Moment of Hg-199}},
\newblock \bibinfo{journal}{Phys. Rev. Lett.} \bibinfo{volume}{102}
  (\bibinfo{year}{2009}) \bibinfo{pages}{101601}.
%Type = Article
\bibitem[{Graner et~al.(2016)Graner, Chen, Lindahl, and Heckel}]{bib:hgupdate}
\bibinfo{author}{B.~Graner}, \bibinfo{author}{Y.~Chen}, \bibinfo{author}{E.~G.
  Lindahl}, \bibinfo{author}{B.~R. Heckel},
\newblock \bibinfo{title}{{Reduced Limit on the Permanent Electric Dipole
  Moment of Hg199}},
\newblock \bibinfo{journal}{Phys. Rev. Lett.} \bibinfo{volume}{116}
  (\bibinfo{year}{2016}) \bibinfo{pages}{161601}. \bibinfo{note}{[Erratum:
  Phys. Rev. Lett.119,no.11,119901(2017)]}.
%Type = Article
\bibitem[{Baker et~al.(2006)}]{bib:baker}
\bibinfo{author}{C.~A. Baker}, et~al.,
\newblock \bibinfo{title}{{An Improved experimental limit on the electric
  dipole moment of the neutron}},
\newblock \bibinfo{journal}{Phys. Rev. Lett.} \bibinfo{volume}{97}
  (\bibinfo{year}{2006}) \bibinfo{pages}{131801}.
%Type = Article
\bibitem[{Pendlebury et~al.(2015)}]{bib:pendlebury}
\bibinfo{author}{J.~M. Pendlebury}, et~al.,
\newblock \bibinfo{title}{{Revised experimental upper limit on the electric
  dipole moment of the neutron}},
\newblock \bibinfo{journal}{Phys. Rev.} \bibinfo{volume}{D92}
  (\bibinfo{year}{2015}) \bibinfo{pages}{092003}.
%Type = Article
\bibitem[{Pendlebury et~al.(2004)}]{bib:gpe1}
\bibinfo{author}{J.~M. Pendlebury}, et~al.,
\newblock \bibinfo{title}{{Geometric-phase-induced false electric dipole moment
  signals for particles in traps}},
\newblock \bibinfo{journal}{Phys. Rev.} \bibinfo{volume}{A70}
  (\bibinfo{year}{2004}) \bibinfo{pages}{032102}.
%Type = Article
\bibitem[{Harris and Pendlebury(2006)}]{bib:gpe2}
\bibinfo{author}{P.~G. Harris}, \bibinfo{author}{J.~M. Pendlebury},
\newblock \bibinfo{title}{{Dipole-field contributions to
  geometric-phase-induced false electric-dipole-moment signals for particles in
  traps}},
\newblock \bibinfo{journal}{Phys. Rev.} \bibinfo{volume}{A73}
  (\bibinfo{year}{2006}) \bibinfo{pages}{014101}.
%Type = Article
\bibitem[{Lamoreaux and Golub(2005)}]{bib:gpe3}
\bibinfo{author}{S.~K. Lamoreaux}, \bibinfo{author}{R.~Golub},
\newblock \bibinfo{title}{{Detailed discussion of a linear electric field
  frequency shift induced in confined gases by a magnetic field gradient:
  Implications for neutron electric-dipole-moment experiments}},
\newblock \bibinfo{journal}{Phys. Rev.} \bibinfo{volume}{A71}
  (\bibinfo{year}{2005}) \bibinfo{pages}{032104}.
%Type = Article
\bibitem[{Barabanov et~al.(2006)Barabanov, Golub, and Lamoreaux}]{bib:gpe4}
\bibinfo{author}{A.~L. Barabanov}, \bibinfo{author}{R.~Golub},
  \bibinfo{author}{S.~K. Lamoreaux},
\newblock \bibinfo{title}{{Linear electric field frequency shift (important for
  next generation electric dipole moment searches) induced in confined gases by
  a magnetic field gradient}},
\newblock \bibinfo{journal}{Phys. Rev.} \bibinfo{volume}{A74}
  (\bibinfo{year}{2006}) \bibinfo{pages}{052115}.
%Type = Article
\bibitem[{Pignol and Roccia(2012)}]{bib:pignol-roccia}
\bibinfo{author}{G.~Pignol}, \bibinfo{author}{S.~Roccia},
\newblock \bibinfo{title}{{Electric dipole moment searches: Reexamination of
  frequency shifts for particles in traps}},
\newblock \bibinfo{journal}{Phys. Rev.} \bibinfo{volume}{A85}
  (\bibinfo{year}{2012}) \bibinfo{pages}{042105}.
%Type = Article
\bibitem[{Afach et~al.(2015)}]{bib:hgcs}
\bibinfo{author}{S.~Afach}, et~al.,
\newblock \bibinfo{title}{{Measurement of a false electric dipole moment signal
  from $^{199}$Hg atoms exposed to an inhomogeneous magnetic field}},
\newblock \bibinfo{journal}{Eur. Phys. J.} \bibinfo{volume}{D69}
  (\bibinfo{year}{2015}) \bibinfo{pages}{225}.
%Type = Phdthesis
\bibitem[{Komposch(2017)}]{bib:komposch}
\bibinfo{author}{S.~Komposch}, \bibinfo{title}{Realization of a
  high-performance laser-based mercury magnetometer for neutron EDM
  experiments}, Ph.D. thesis, ETH Z\"urich, \bibinfo{year}{2017}.
%Type = Book
\bibitem[{Jackson(1975)}]{bib:jackson}
\bibinfo{author}{J.~D. Jackson}, \bibinfo{title}{Classical Electrodynamics},
  \bibinfo{edition}{2nd} ed., \bibinfo{publisher}{Wiley}, \bibinfo{address}{New
  York, {NY}}, \bibinfo{year}{1975}.
%Type = Book
\bibitem[{Sakurai(1994)}]{bib:sakurai}
\bibinfo{author}{J.~J. Sakurai}, \bibinfo{title}{{Modern Quantum Mechanics,
  rev. ed.}}, \bibinfo{publisher}{Addison-Wesley}, \bibinfo{address}{Reading,
  MA}, \bibinfo{year}{1994}.
%Type = Misc
\bibitem[{Abel et~al.(2018)}]{bib:pignol-priv}
\bibinfo{author}{C.~Abel}, et~al., \bibinfo{title}{{Magnetic field uniformity
  in neutron electric dipole moment experiments}},
  \bibinfo{howpublished}{{arXiv:1811.06085 [physics.ins-det]}},
  \bibinfo{year}{2018}.

\end{thebibliography}

%% Authors are advised to submit their bibtex database files. They are
%% requested to list a bibtex style file in the manuscript if they do
%% not want to use model1-num-names.bst.

%% References without bibTeX database:

% \begin{thebibliography}{00}

%% \bibitem must have the following form:
%%   \bibitem{key}...
%%

% \bibitem{}

% \end{thebibliography}

\end{document}